\documentclass{svproc}
\usepackage{graphicx}%
\usepackage{multirow}%
\usepackage{amsmath,amssymb,amsfonts}%
\usepackage{mathrsfs}%
\usepackage[title]{appendix}%
\usepackage{xcolor}%
\usepackage{textcomp}%
\usepackage{manyfoot}%
\usepackage{booktabs}%
\usepackage{listings}%
\usepackage{url}
\usepackage{cite}

\usepackage[ruled,lined]{algorithm2e}

\usepackage{caption}
\usepackage{amsfonts}
\usepackage{hyperref}
\usepackage{easyReview}

\begin{document}

%
\title{INDoRI: Indian Dataset of Recipes and Ingredients and its Ingredient Network}
\titlerunning{INDoRI and its Ingredient Network}  
%
\author{Sandeep Khanna\inst{1}, Chiranjoy Chattopadhyay\inst{2},
	Suman Kundu\inst{1}}
\authorrunning{Khanna et al.} 
%
\tocauthor{}
\institute{Indian Institute of Technology Jodhpur,\\
	\email{\{khanna.1,suman\}@iitj.ac.in}
	\and
	FLAME University, Pune\\
        \email{chiranjoy.chattopadhyay@flame.edu.in}}

\maketitle   

\begin{abstract}
Exploring and comprehending the culinary heritage of a nation holds a captivating allure. It offers insights into the structure and qualities of its cuisine. The endeavor becomes more accessible with the availability of a well-organized dataset. In this paper, we present the introduction of INDoRI (Indian Dataset of Recipes and Ingredients), a compilation drawn from seven distinct online platforms, representing 18 regions within the Indian subcontinent. This comprehensive geographical span ensures a portrayal of the rich variety within culinary practices. Furthermore, we introduce a unique collection of stop words, referred to as ISW (Ingredient Stop Words), manually tuned for the culinary domain. We assess the validity of ISW in the context of global cuisines beyond Indian culinary tradition. Subsequently, an ingredient network (InN) is constructed, highlighting interconnections among ingredients sourced from different recipes. We delve into both the defining attributes of INDoRI and the communal dimensions of InN. Additionally, we outline the potential applications that can be developed leveraging this dataset. Addressing one of the applications, we demonstrated a research problem on InN with a simple weighted community detection algorithm. Furthermore, we provide a comparative analysis of the results obtained with this algorithm against those generated by two baselines.

\end{abstract}

\section{Introduction}

India, characterized by its rich tapestry of cultures, hosts a plethora of distinct cuisines. Tackling food computing challenges within this culinary landscape is indeed complex. One significant hurdle stems from the dearth of structured data that spans India's diverse cuisines despite numerous websites house extensive recipe databases. The reason for the same is that the information available therein is predominantly unstructured, comprising text and multimedia content.

This paper introduces the Indian Dataset of Recipes and Ingredients (INDoRI), encompassing a total of $5187$ recipes. Recipes were extracted and gathered from different online platforms \cite{iff:58,eir:57,vrec:56,hrec:54,allrec:55,san:58}. These recipes span a variety of Indian cuisines, reflecting the rich cultural diversity across regions such as Punjabi, Bengali, and Gujarati. INDoRI stands as a structured repository of recipes and their corresponding ingredients. Further, the dataset includes a graph-based representation of ingredient relationships, namely, ingredient network (InN). InN is formed by capturing ingredient relationships based on their co-occurrence within recipes. 

Extracting meaningful information from widely available recipes from the web, required to remove several stop words apart from the natural language stop words. For instance, terms like ``pinch" and ``mix" appear with the list of ingredient in a recipe needs to be removed to extract actual ingredient. We introduced a novel set of $572$ stop words aligning with food ingredients and named that set as Ingredient Stop Words (ISW). Furthermore, validity of ISW is checked with three other cuisines i.e., Japanese, American and Italian. The use of these stop words proves instrumental in effectively extracting and refining ingredient names. 

In summary, the paper presents

\begin{enumerate}
    \item Proposal of INDoRI, a dataset of Recipes and Ingredients of Indian cuisines. It includes over 5K recipes with 18 different cuisines. The characteristics and possible applications of the data set are reported.
    \item A novel set of stop words ISW for the culinary domain. 
    \item Construction of the Ingredient Network (InN) on top of INDoRI.
    \item Demonstrated a research problem on InN with a simple weighted community detection algorithm (WABCD).
\end{enumerate}

\section{Literature Survey}
\label{sec:ls}
\paragraph{\textbf{Datasets:}}
Over the course of time, numerous benchmark food datasets have been introduced in research literature. For instance, Matsuda et al. \cite{Matsuda:46} introduced a Japanese food image dataset in 2012, encompassing a collection of 14,361 images. In 2014, Bossard et al. \cite{Lukas:48} released the ETHZ Food-101 dataset. The year 2016 saw the unveiling of a large dataset by Rich et al. \cite{Rich:47} containing 800 thousand images. Many of these existing datasets are focused on images, although a few exceptions exist in the form of datasets oriented towards recipes. Notably, three recipe-centric datasets emerged in 2018. These are: a recipe question-answering dataset by Semih et al. \cite{Semih:50}, comprising approximately 36k questions that users can query against the dataset; Epic Kitchen, introduced by Damen et al. \cite{Damen:51}, featuring cooking videos and accompanying recipes; and the extensive ``Recipe1M" dataset containing both recipes and images, brought forth by Salvador et al. \cite{Amaia:49}. The work of Salvador et al. \cite{Amaia:49} notably focuses on embedding recipes and images. Furthermore, they extended their dataset to create ``Recipe1M+" \cite{Javier:59}.

\paragraph{\textbf{Ingredient Network:}}
Over time, researchers have explored ingredient networks in various contexts. One such study \cite{teng2012recipe} resulted in the creation of two ingredient networks: ``complement" and ``substitute". The complement network exhibited two distinct communities, one centered around savory ingredients and the other around sweet ingredients. On the other hand, the substitute network was constructed based on user-generated suggestions, offering alternative ingredient choices for specific recipes. Similarly, another work \cite{nyati2021characterize} focused on two types of networks: ingredient-ingredient and recipe-ingredient networks. These networks were designed to recommend recipes to users based on the ingredients they had available. By analyzing the relationships between ingredients and recipes, the system could suggest suitable recipes that aligned with the user's resources. Apart from recipe recommendations, ingredient networks have also been applied to food recognition tasks. For instance, Min et al. \cite{min2019ingredient} achieved food recognition by developing an innovative Ingredient-Guided Cascaded Multi-Attention Network. This approach utilized the ingredient network to enhance the accuracy of food recognition systems, leveraging the knowledge of the associations among the food ingredients. However, we introduced INDoRI, which distinguishes itself by encompassing not only recipes, ingredients, and cooking instructions, but also comprehensive cuisine information representative of the entirety of India.

\section{Indian Dataset of Recipes and Ingredients (INDoRI)}
\label{sec:INDoRInew}
Creating a comprehensive dataset of Indian cuisines possesses unique challenges. One of them is to compiling recipes that span diverse cultural landscape of India. Due to the same reason one may not find all the recipes from one single web portal. As there is no common data format available, each portal present data differently and the data are unstructured. Hence the second challenge is to extract meaningful information from it. We consider seven different recipe websites to address the first challenge. All the unstructured data therein are crawled using Python script. Basic cleaning is performed on the collected data and the following methodology is used to structurized it using both tabular and network structures.

\paragraph{\textbf{Identification of novel stop words for food ingredients (ISW):}}
Amidst the data preparation phase, novel food-related stop words were introduced. Notable examples encompass `kg,' `gms,' `cup,' `tbls,' `pinch,' `chopped,' `boiled,' `sliced,' and `split'. $527$ specific keywords are identified, scrutinized, and extracted manually from the ingredient data. The validity of ISW was tested on other global cuisines, including Japanese \cite{Jap}, Italian \cite{Ita} and American \cite{Ame}. For each
of the cuisine hundred recipes were taken along with the ingredients needed to prepare them. The ingredient names were extracted manually and through stop word removal using ISW. The details were reported in Table \ref{Tab:res}. A comprehensive breakdown of the calculations and results are presented in an online repository \footnote{Link to the supplementary material: \url{ https://shorturl.at/gwzFN}}. 


\begin{table}[!htbp]
\centering
\caption{Cuisine wise accuracy statistics: ISW accuracy for global Cuisine.}
\label{Tab:res}
\begin{tabular}{|c|c|c|c|}
\hline
\textbf{Cuisine} & \textbf{Avg. Accuracy} & \textbf{Min. Accuracy} & \textbf{Max. Accuracy} \\ \hline
\textbf{Indian}           & 81.98                  & 80.0                   & 92.85                  \\ \hline
\textbf{Italian}          & 75.42                  & 68.75                  & 83.33                  \\ \hline
\textbf{Japanese}         & 53.36                  & 42.85                  & 60.0                   \\ \hline
\textbf{American}         & 72.31                  & 62.50                  & 85.71                   \\ \hline
\end{tabular}
\end{table}

\paragraph{\textbf{Removal of stop words and numbers:}}
The exclusion of stop words and numerical values from ingredients yielded beneficial results in obtaining clean ingredient names.  Solely ingredient names are employed to construct the ingredient network (InN). Nonetheless, these numerical values can potentially be considered for recommendation purposes hence kept separately.

\begin{figure}[t]
    \centering
    \includegraphics[width=\textwidth]{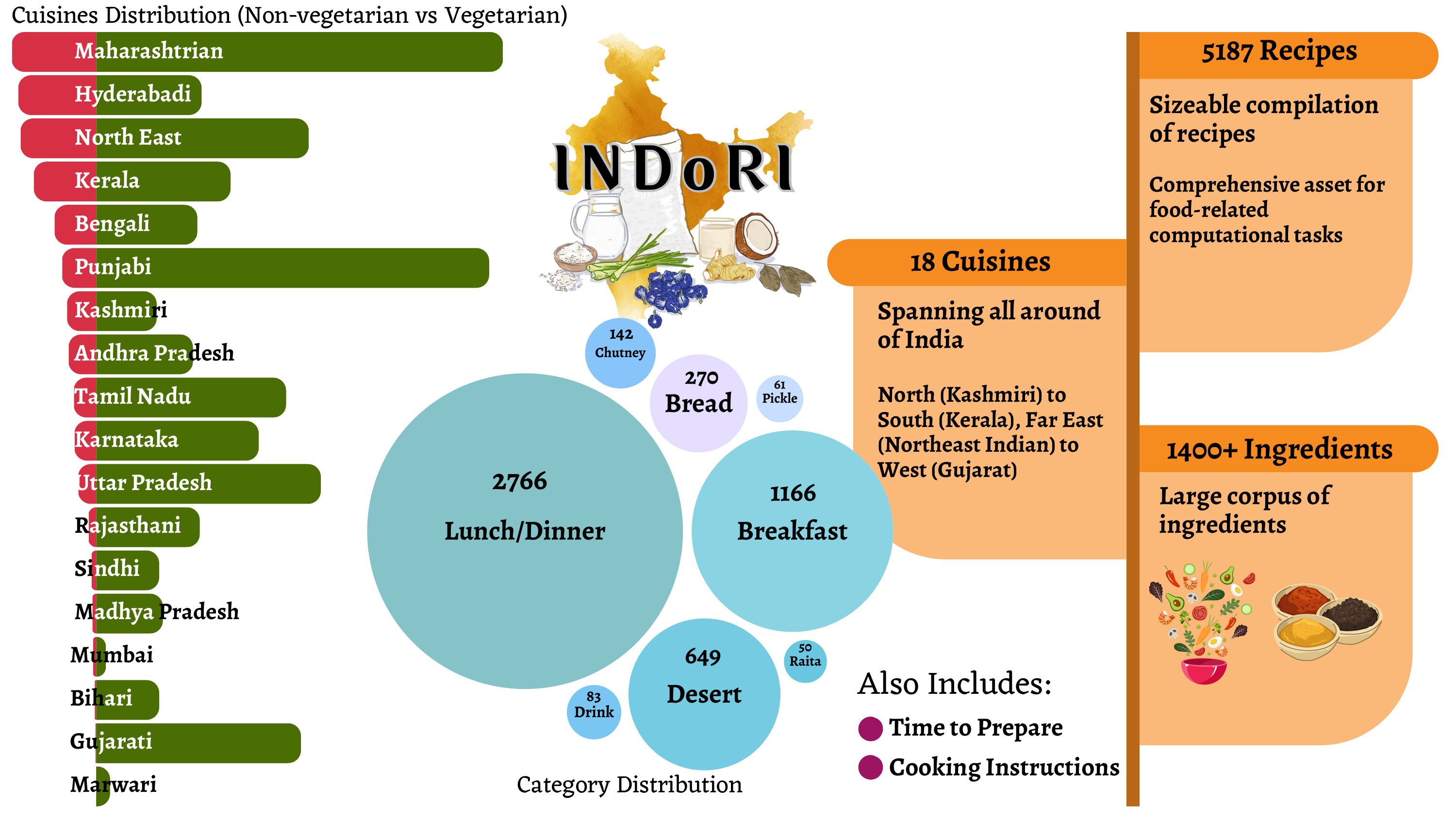}
    \caption{Key Characteristics of INDoRI}
    \label{fig:enter-label}
\end{figure}

\paragraph{\textbf{Characteristics of INDoRI:}}
INDoRI stands as a unique and innovative Indian recipes dataset, distinguishing itself from conventional counterparts. It contains a total of 5187 recipe, presenting a diverse array of culinary offerings. Additionally INDoRI encompasses additional attributes such as cuisine, category, and preparation time. All recipes are classified into 8 different types. Apart from 925 unclassified recipes rest are also categorized into 18 different cuisines. Fig. \ref{fig:enter-label} shows the key characteristics of INDoRI. In order to examine the interrelationships among ingredients, we formed a network of ingredients referred to as the Ingredient Network (InN). Further information regarding this network is outlined in the subsequent Section.

\begin{table}[!h]
	\begin{minipage}{0.5\linewidth}
		\label{table:student}
		\centering
		\begin{tabular}{ll}
\hline

\multicolumn{2}{l}{\textbf{Statistics}} \\ \hline
Directed                                     & No                   \\ \hline
Weighted                                     & Yes                  \\ \hline
Nodes                                        & 1433                 \\ \hline
Edges                                        & 30464               \\ \hline
Average Clustering Coefficient               & 0.8455               \\ \hline
Number of Triangles                          & 424048               \\ \hline
Fraction of Closed Triangles                 & 0.3485               \\ \hline
Diameter                                     & 4                    \\ \hline
Average Edge Weight                          & 39.7861              \\ \hline
\end{tabular}
	\end{minipage}\hfill
	\begin{minipage}{0.50\linewidth}
		\centering
		\includegraphics[width=4.5cm, height=4.1cm]{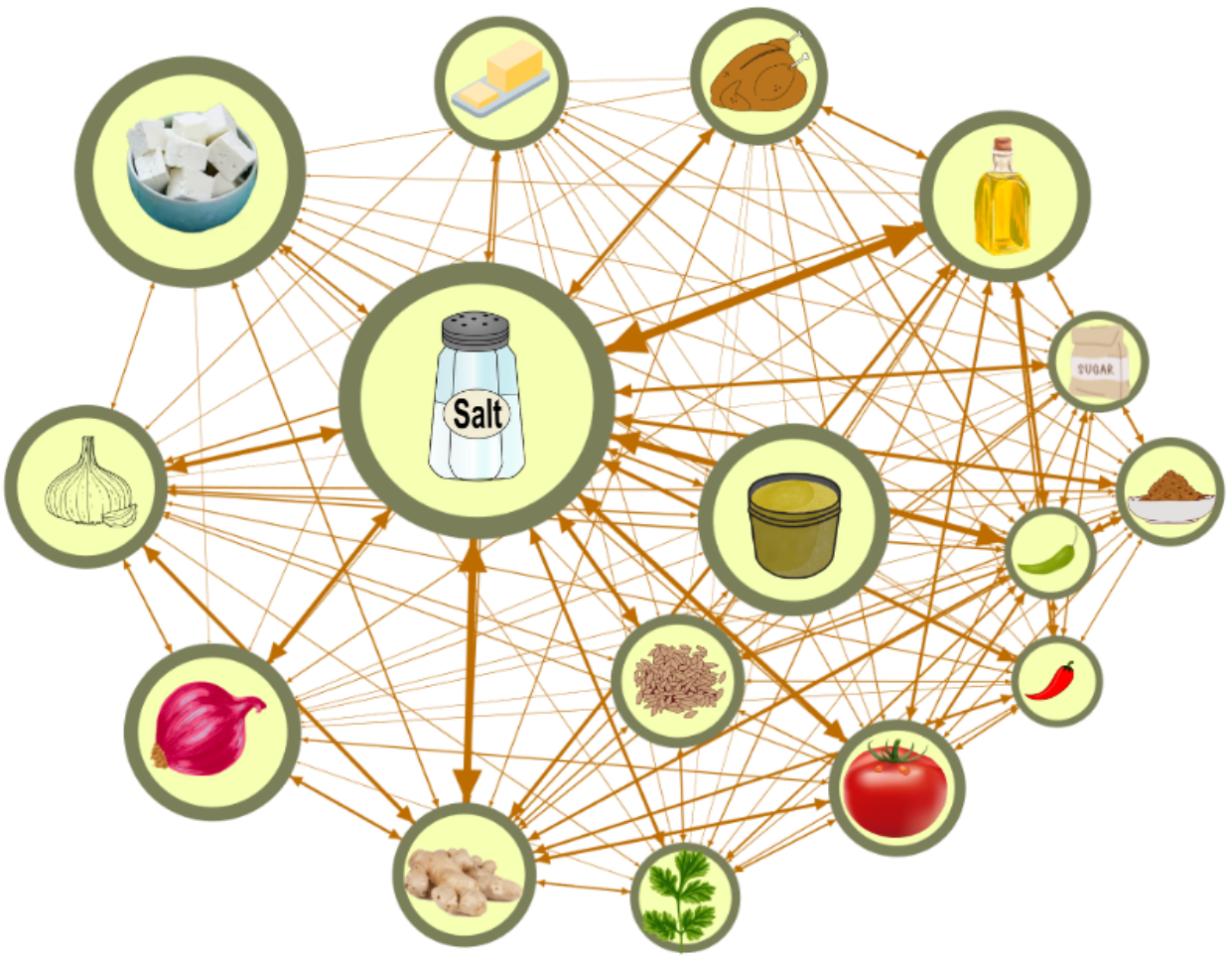}
	\end{minipage}\hfill

        \begin{minipage}{0.3\linewidth}
		\centering
		\includegraphics[width=6.5cm, height=6.1cm]{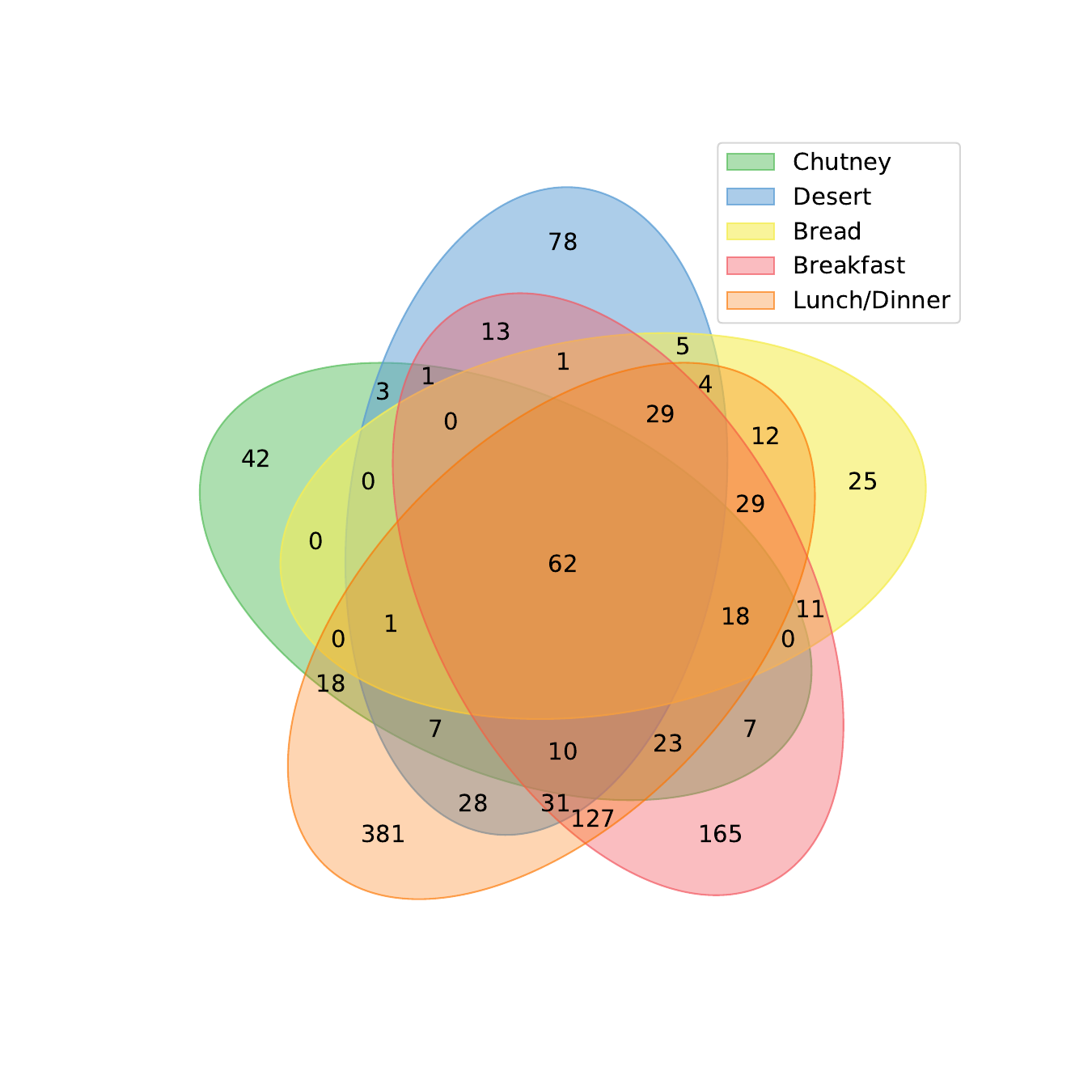}
	\end{minipage}\hfill
         \begin{minipage}{0.55\linewidth}
		\centering
		\includegraphics[width=6.5cm, height=6.1cm]{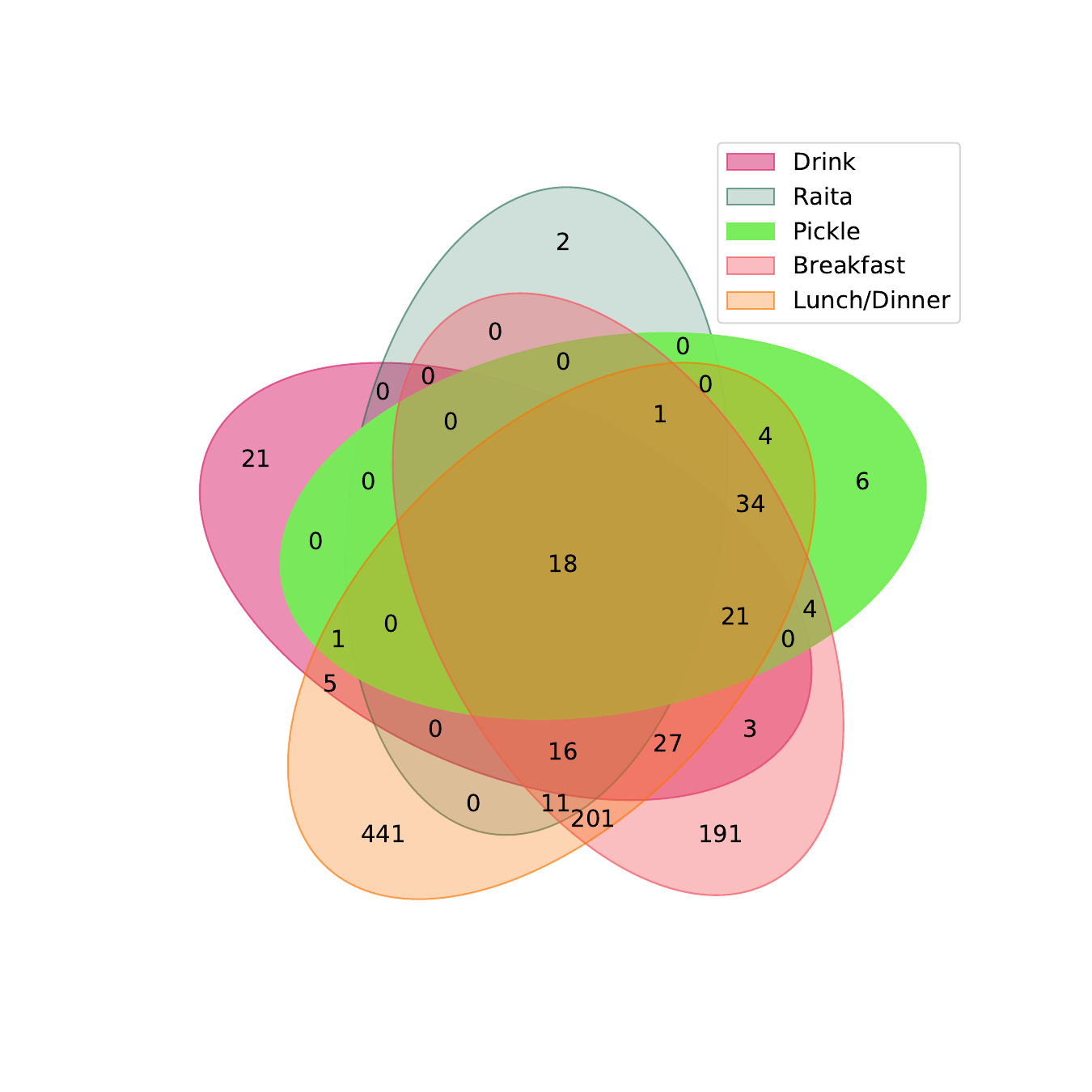}
	\end{minipage}
\captionof{figure}{Characteristics and statistics of InN and ingredients.}
\label{fig:inncha}
\end{table}

\subsection{Ingredient Network Construction}
We constructed the ingredient network out of INDoRI where each node is an ingredient. A link is constructed when two ingredient appear in the same recipe. Total of 30,464 relationships were found among all ingredients. The ingredient network is a graph $G (V,E,w)$, where $V$ is a set of ingredients, $E$ is the connections between ingredients and $w: V \times V \rightarrow \mathbb{R}$ of an edge signifies the number of association between ingredients in different recipes. The more they appear together in diverse recipes, more stronger is the association. The strongest association, is between salt and oil, appearing together in 1958 recipes. 

\begin{figure}[!htbp]
\center
    \begin{minipage}[b]{.45\linewidth}
        \includegraphics[width=\textwidth]{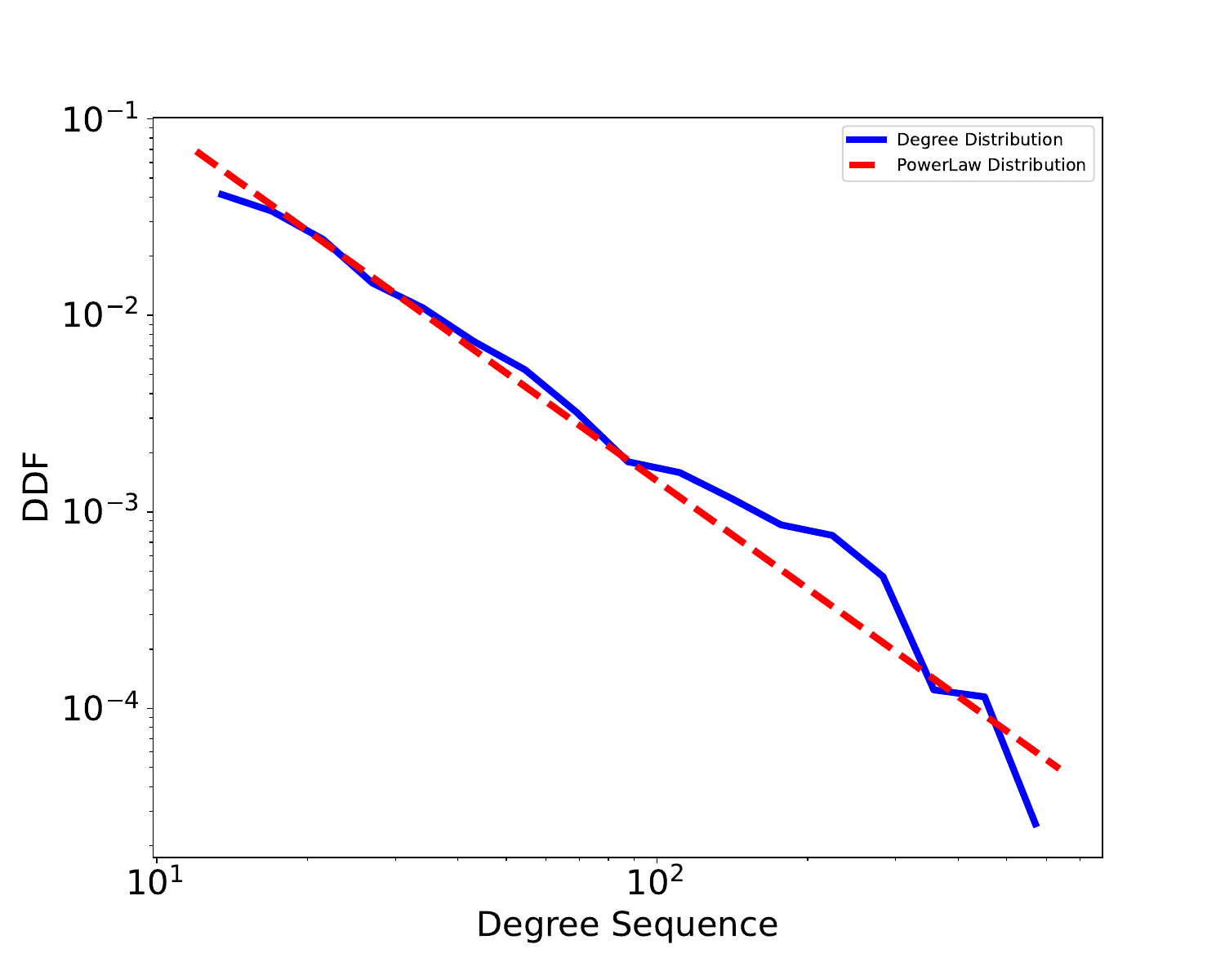}
        \centerline{\small{(a) Degree Distribution}}\medskip
    \end{minipage}
    \begin{minipage}[b]{0.45\linewidth}
        \includegraphics[width=\textwidth]{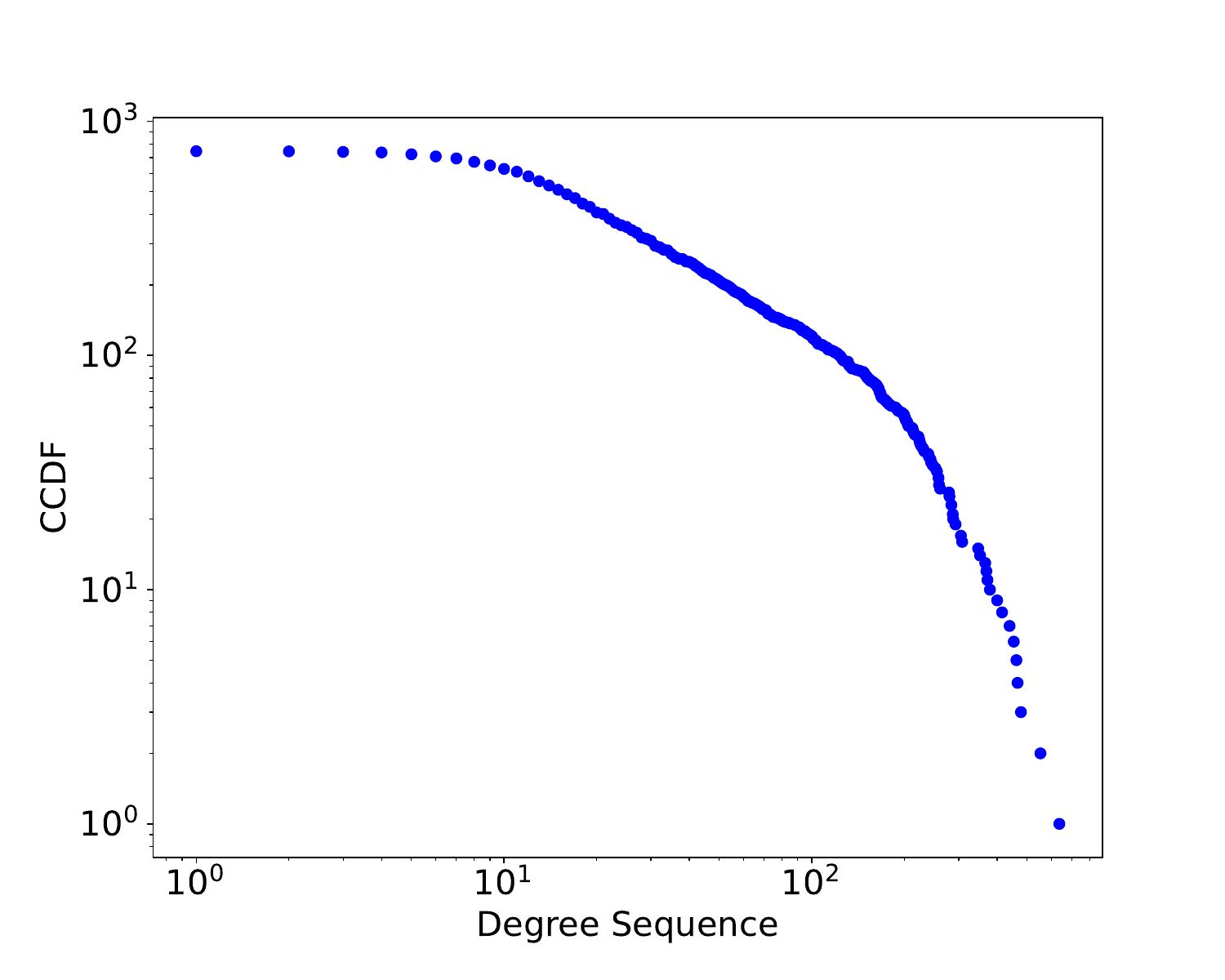}
        \centerline{\small{(b) Cumulative Degree Distribution}}\medskip
        \label{fig:dcbb}
    \end{minipage}
    \caption{Distribution Plot of InN. }
\label{fig:degdist}
\end{figure}

\paragraph{\textbf{Characteristics of ingredients and InN}}
Sample sub graph of InN is shown in top right of Fig. \ref{fig:inncha}. Here thick edges represent stronger associations, while thinner edges represent weaker associations. The size of the node shows the degree. The bigger the size greater is the degree. Top left Table shows the statistics of the network InN. 
We also investigated the presence of ingredients in multiple recipe categories. The bottom images of Fig. \ref{fig:inncha} represents the ingredient overlaps. While the left image provide overlap across five recipe categories viz Chutney, Desert, Bread, Breakfast and Lunch\slash Dinner the right image shows the overlap among Drink, Raita, Pickle, Breakfast and Lunch\slash Dinner. It is evident that there were 62 ingredients shared among all categories in the left image and overlap of 18 ingredients is found in the right image. This gives an interesting observation of the distinct communities for each category. The degree distribution of InN follows a power law, making it a scale-free network as shown in Figs. \ref{fig:degdist} (a) and (b) shows the cumulative degree distribution.  

\subsection{Communities in InN}
\label{sec:chal}
The average clustering coefficient of InN is measured as 0.8455, indicating a higher tendency for nodes to form clusters or groups within the network. We employed the weighted Leiden algorithm \cite{traag2019louvain} to identify community structure of InN. The outcomes are presented in Section \ref{sec:qa}, highlighting that the network is partitioned into five distinct communities. We tried to uncover the inherent characteristics of each partition, leading us to recognize a distinct pattern. Specifically, we observed that the majority of categories, excluding dessert, are having strong associations with the first partition or community. Conversely, the exceptions displayed associations with the second partition. This observation presents a fascinating challenge for researchers to devise a weighted algorithm tailored for community detection within InN. Such an algorithm has the potential to identify diverse trends in the network structure.

\begin{table}[!htbp]
\centering
\caption{Potential Applications on INDoRI and InN.}
\resizebox{\columnwidth}{!}{%
\begin{tabular}{ll}
\hline
\multicolumn{2}{c}{\textbf{INDoRI}} \\ \hline
\textbf{Application} &
  \textbf{Description} \\ \hline
Recipe Categorization &
  \begin{tabular}[c]{@{}l@{}}Automatic categorization of recipes into categories such as breakfast, \\ lunch, dinner etc. based on text descriptions like ingredients and \\ cooking instructions.\end{tabular} \\ \hline
Cuisine Classification &
  \begin{tabular}[c]{@{}l@{}}Automatic categorization of recipes into cuisines \\ such as Punjabi, Bengali, Hyderabadi etc. based on text descriptions\\ like ingredients and cooking instructions.\end{tabular} \\ \hline
\multicolumn{2}{c}{\textbf{InN}} \\ \hline
\textbf{Application} &
  \textbf{Description} \\ \hline
Community Identification &
  \begin{tabular}[c]{@{}l@{}}Algorithms to identify communities in the Ingredient Network (InN) \\ where each community can be correlated with cuisine or category.\end{tabular} \\ \hline
Ingredient Pair Prediction &
  \begin{tabular}[c]{@{}l@{}}Development of methods to predict occurring pairs of ingredients \\ in recipes, aiding link prediction and recommendation.\end{tabular} \\ \hline
\multicolumn{2}{c}{\textbf{INDoRI + InN}} \\ \hline
\textbf{Application} &
  \textbf{Description} \\ \hline
Recipe Similarity &
  \begin{tabular}[c]{@{}l@{}}Design of techniques that measure recipe similarity or dissimilarity \\ based on ingredient overlap, cooking techniques, and other attributes.\end{tabular} \\ \hline
Ingredients based Recipe Recommendation &
  \begin{tabular}[c]{@{}l@{}}Proposal of recommendation algorithms predicting recipes based on \\ ingredient availability, offering personalized suggestions.\end{tabular} \\ \hline
\end{tabular}%
}
\label{Tab:app}
\end{table}

\section{Applications on INDoRI and InN}
\label{sec:app}
Food Computing is defined as the study of food and its properties using computational methods and methodologies \cite{Min:52}. One such method is modeling and simulation. It involves many tasks such as acquiring, analyzing, recognition \cite{xu2015geolocalized}, recommendation of food and recipes. Considering the characteristics of this dataset, researchers have the opportunity to delve into the tasks both on INDoRI and InN. Some of the potential applications are listed in Table \ref{Tab:app}. The outcomes of two applications of INDoRI are described in the online repository \footnote{Link to the supplementary material: \url{ https://shorturl.at/gwzFN}}. 

\begin{algorithm}[!hbtp]
\caption{WABCD $(G, V, E, w)$} \label{alg:new}
\footnotesize
\KwIn{$G$: InN graph, $V$: set of vertices of $G$, $E$: set of edges of $G$ and $w$: set of weights on edges}
\KwOut{Acquired communities in \textit{dictnew}}
\nl \textit{dictnew} = \{\}\;            \label{WABCD:1}
\nl \For{i in range (0, len($V$))}{
\nl     \textit{dictnew}[$V[i]$] = $V[i]$\;                                                         \label{WABCD:2} 
    }
\nl \While{True}{                        \label{WABCD:3}                 
    \nl \textit{dict} = dictnew.copy()\;   \label{WABCD:4}
    \nl \For{key1 in list(dict.keys())}{   \label{WABCD:5}
    \nl     \textit{bestinc} = 0; \textit{c} = 0; \textit{key} = \textit{dict}[\textit{key1}]\; 
    \nl     \For{key2 in list(dict.keys())}{     \label{WABCD:6} 
    \nl          \textit{newkey} = \textit{dict}[\textit{key2}]\; 
    \nl          \If {key1!= key2 and len(key) $>$ 0 and len(newkey) $>$ 0}{ \label{WABCD:7}
    \nl             \textit{sumweight} = 0\;  
    \nl             \For {m in key}{
    \nl                 \For {n in newkey}{
    \nl                     \If {$G$ hasedge(m,n)}{ \label{WABCD:8}
    \nl                         \textit{sumweight} = \textit{sumweight} + $G$.getedgedata(m, n)[weight]; \label{WABCD:9}\textit{c}+=1\;
            }
    \nl                     \Else{
    \nl                           continue\; \label{WABCD:10}
                            }
                        }
                    } 
    \nl             \If {c $>$ 0}{\textit{sumweight} = \textit{sumweight} / \textit{c}\;        
                    }
    \nl             \textit{accnode} = \textit{sumweight}\; 
    \nl             \If {accnode $>$ bestinc}{
    \nl                 \textit{bestinc} = \textit{accnode}; k = \textit{newkey}; q = \textit{key}\;
                    }
                }
            }
    \nl     \If{bestinc $>$ 0}{         \label{WABCD:11}
    \nl         \If{dictnew[key1] != -1}{
    \nl             \For{qw in k}{
    \nl                 \textit{dictnew}[\textit{key1}].append(\textit{qw})\; \label{WABCD:12}
                    }
    \nl             \textit{dictnew}[\textit{key1}] = -1\;
                } \label{WABCD:13}
    \nl         delete \textit{dict}[\textit{key1}]\;
            }
        }   \label{WABCD:16}
    \nl \For{key in list(dictnew.keys())}{
    \nl      \If{dictnew[key] == -1}{
    \nl          \textit{dictnew}.pop(\textit{key})\;
              } \label{WABCD:14}
        }
    \nl \If{bestinc == 0}{   \label{WABCD:15}                                  
    \nl     break\;   \label{WABCD:17}
        } }   
\end{algorithm}

\subsection{Example: Community Detection for better categorization of ingredients}
\label{sec:qa}
Addressing the challenge we have discussed in Section \ref{sec:chal} for community identification in InN, we proposed a simple Weighted Association Based Community Detection (WABCD) algorithm that groups nodes based on the strong association between them. The input to the algorithm is a weighted graph $G(V,E,w)$ and it outputs community structure therein. 
The algorithm (Algorithm \ref{alg:new}) works in the following manner. At the outset, every vertex denotes a unique community. In the first cycle, communities merge according to the most substantial weighted edge between two vertices. Starting from the second iteration, each vertex within a community is compared with vertices from other communities, and the average weight between the communities is calculated. Merging occurs based on the highest average value between two communities. The algorithm terminates when the average weight computed in an iteration is lower than that calculated in the previous iteration.  

\begin{figure*}[!htbp]
\centering
\footnotesize
    \includegraphics[width=0.8\textwidth]{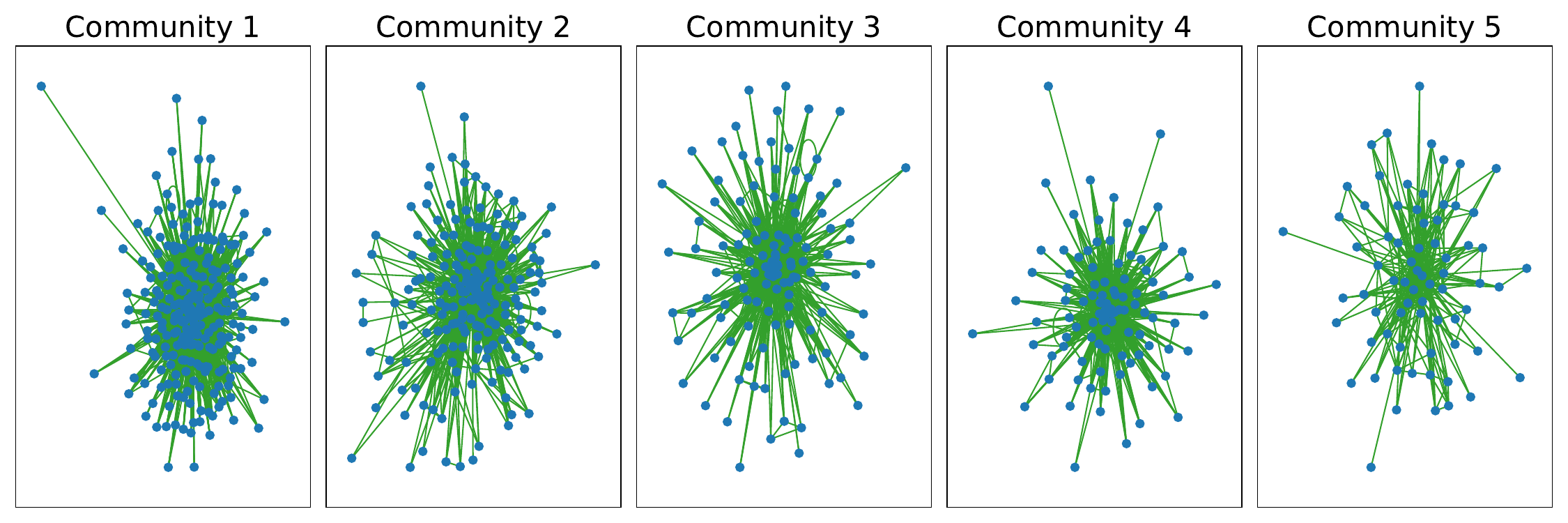}
    \centerline{\small{(a) Weighted Leiden}}
    \includegraphics[width=0.8\textwidth]{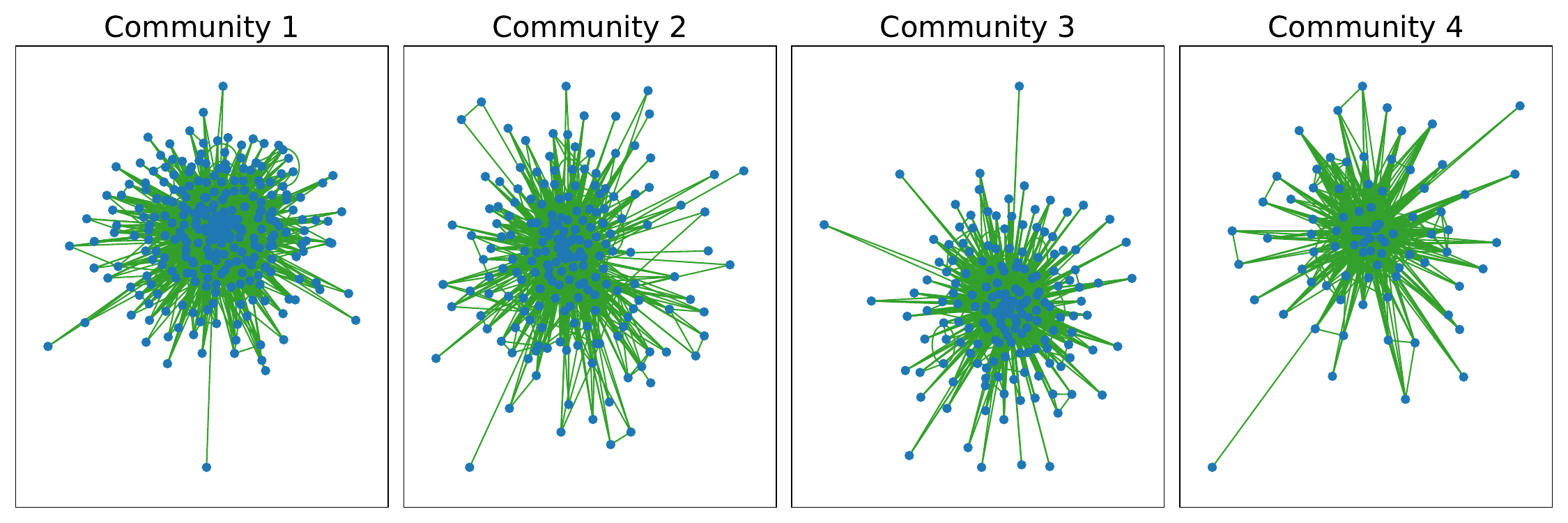}
    \centerline{\small{(b) Weighted Louvain}}
    \includegraphics[width=0.8\textwidth]{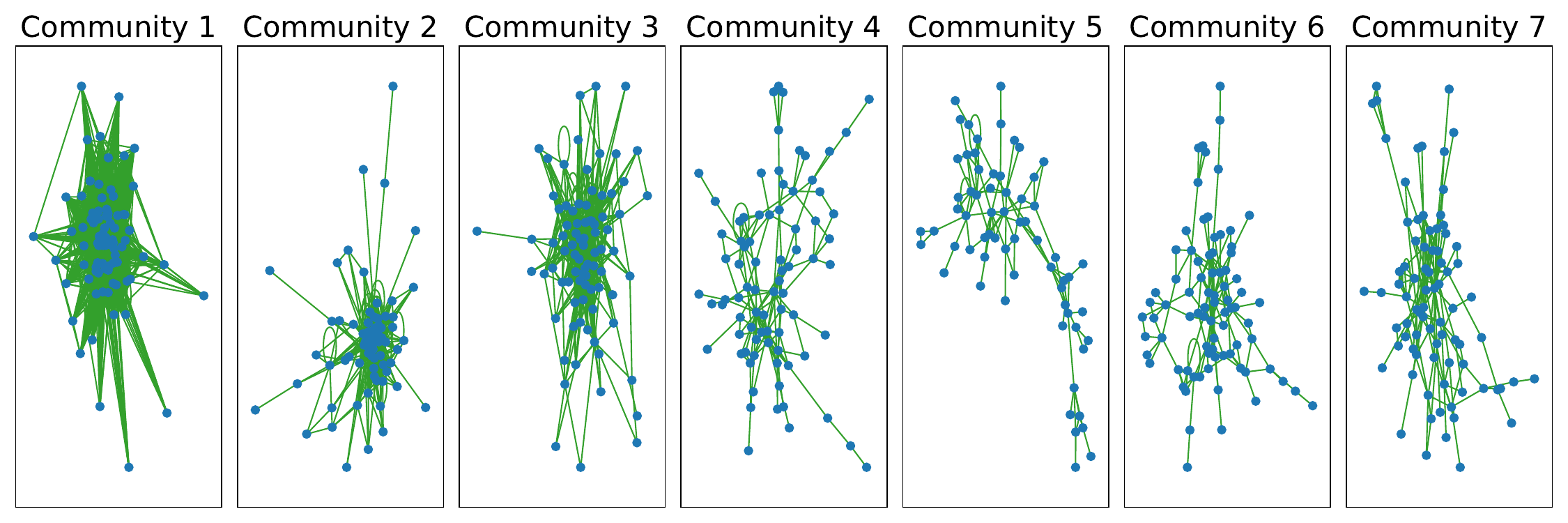}
    \centerline{\small{(c) WABCD}}
    \caption{Results from Different Community Detection Algorithms a) Weighted Leiden detects $5$ communities b) Weighted Louvain detects $4$ communities c) WABCD detects $7$ communities 
    }
\label{fig:foolbar}    
\end{figure*}

\begin{table}[!htbp]
\centering
\footnotesize
\caption{Comparison of community detection algorithms}
\begin{tabular}{|c|c|c|c|}
\hline
\textbf{}   & \textbf{Weighted Leiden} & \textbf{Weighted Louvain} & \textbf{WABCD} \\ \hline
\textbf{C1} & Lunch/Dinner Recipes   & Lunch/Dinner Recipes    & Bread Recipes          \\ \hline
\textbf{C2} & Desert Recipes   & Desert Recipes    & Bread Recipes          \\ \hline
\textbf{C3} & Lunch/Dinner Recipes   & Lunch/Dinner Recipes    & Lunch/Dinner Recipes       \\ \hline
\textbf{C4} & Lunch/Dinner Recipes   & Lunch/Dinner Recipes    & Drink Recipes     \\ \hline
\textbf{C5} & Lunch/Dinner Recipes   &      -            & Lunch/Dinner Recipes            \\ \hline
\textbf{C6} &      -           &     -             & Dessert Recipes          \\ \hline
\textbf{C7} &       -          &      -            & Lunch/Dinner Recipes                \\ \hline
\end{tabular}
\label{Table:comp}
\end{table}

\paragraph{\textbf{Comparison of WABCD with baselines}}
\label{comp}
We compare the proposed WABCD algorithm with other community detection algorithms. Baseline algorithms considered were weighted Leiden and weighted Louvain \cite{blondel2008fast}. The results were shown in  Fig. \ref{fig:foolbar}. The communities identified by weighed Leiden, Louvain and WABCD is 5, 4, 7 respectively. To uncover the inherent characteristics of each partition we have created multiple sub-graphs based on the category of recipes and compare them with the communities obtained from all three algorithms. The results were shown in Table \ref{Table:comp}. One may observed that with both weighted Leiden  and Louvain algorithm, the second community exhibit connection with recipe category Desert whereas the rest tend to have more association with Lunch/Dinner category. Conversely, the WABCD approach succeeds in identifying four prominent recipe categories: Bread, Lunch/Dinner, Drink, and Deserts. However the desired number of communities is 8 with overlap in between as shown in Fig \ref{fig:inncha} and the problem remain open to solve.

\section{Conclusion}
\label{sec:conc}
This paper presented our INDoRI dataset with a general characterization along with its ingredient network. We thoroughly examined and shown its distinctive features and attributes. Furthermore, we have put forth a set of novel stop words specifically tailored for the food ingredients. The creation of the Ingredient network (InN) from ingredient interconnections has been a focal point, with a comprehensive analysis on community identification. Our discourse extends to addressing the potential applications on top of INDoRI and InN. We present and compare the communities identified using WABCD and other baseline community detection algorithms. Overall, INDoRI and InN not only enriches our understanding of Indian cuisine but also opens up fresh avenues for research, encouraging a deeper exploration of its culinary intricacies.

\bibliographystyle{ieeetr}
\bibliography{sample-base}

\end{document}